\documentclass[10pt,aps,prl,floatfix,superscriptaddress,twocolumn,footinbib]{revtex4-2}
\usepackage{graphicx}
\usepackage{bm}

\usepackage{amsmath, amssymb, amsfonts}

\usepackage{booktabs}
\usepackage{xcolor}

\begin{document}

\title{Realizing Two-Dimensional Altermagnets with 100\% Spin Polarization along the [110] Direction in Ambidentate-Ligand Based Square Organometallic Lattices}

\author{Qing-Bo Liu}
\affiliation{Hubei Key Laboratory of Optical Information and Pattern Recognition, School of Optical Information and Energy Engineering, School of Mathematics and Physics, Wuhan Institute of Technology, Wuhan, 430073, China}
\affiliation{Key Laboratory of Precision and Intelligent Chemistry, University of Science and Technology of China, Hefei, Anhui 230026, China}

\author{Xin-Yi Chen}
\affiliation{Hubei Key Laboratory of Optical Information and Pattern Recognition, School of Optical Information and Energy Engineering, School of Mathematics and Physics, Wuhan Institute of Technology, Wuhan, 430073, China}

\author{Xiangyang Li}
\altaffiliation{xyangli@ustc.edu.cn}
\affiliation{Hefei National Laboratory, University of Science and Technology of China, Hefei, Anhui 230088, China}

\author{Xingxing Li}
\altaffiliation{lixx@ustc.edu.cn}
\affiliation{Key Laboratory of Precision and Intelligent Chemistry, University of Science and Technology of China, Hefei, Anhui 230026, China}
\affiliation{Hefei National Laboratory, University of Science and Technology of China, Hefei, Anhui 230088, China}

\author{Jinlong Yang}
\altaffiliation{jlyang@ustc.edu.cn}
\affiliation{Key Laboratory of Precision and Intelligent Chemistry, University of Science and Technology of China, Hefei, Anhui 230026, China}
\affiliation{Hefei National Laboratory, University of Science and Technology of China, Hefei, Anhui 230088, China}
\date{\today}

\begin{abstract}

Altermagnetism, as a novel magnetic paradigm, not only possesses a zero net magnetic moment, but also exhibits a substantial nonrelativistic spin-splitting effect. Currently, developing a simple and experimentally feasible method for designing altermagnets is eagerly pursued. Moreover, achieving 100\% spin polarization along certain directions in valence or conduction bands for altermagnets remains challenging. To address these issues, based on a tight-binding model, we propose a viable chemical strategy for constructing organometallic altermagnets by using ambidentate ligands. Furthermore, in these organometallic altermagnets, direction-dependent 100\% spin polarization along the [110] direction in band edges can be induced via the coexistence of altermagnetically spin-split dispersive and flat bands, where the flat band suppresses electron transport in one spin channel and the dispersive band enables conduction in the other spin channel. Then, via first-principles calculations, a screening of 90 two-dimensional ambidentate-ligand based square organometallic lattices leads to the identification of three altermagnetic materials, namely Fe(taz)$_2$Cl$_2$ (taz = 1,2,4-triazine), Co(app)$_2$Cl$_2$, and Fe(app)$_2$Cl$_2$ (app = 1,4-azaphosphinine), all of which possess 100\% spin polarization along the [110] direction in valence bands, as well as notable piezomagnetic behavior. This study paves the way for achieving direction-dependent fully spin-polarization in altermagnetic materials.

\end{abstract}

\maketitle

{\bf{1. Introduction}}

Altermagnetism, an emerging class of collinear magnetic order characterized by vanishing net magnetization and momentum-dependent spin splitting {\color{blue}\cite{1,2,3,4,5}},
has opened a new frontier in spintronics.
Unlike conventional ferro- or antiferromagnetism,
altermagnetic materials exhibit spin-polarized electronic bands protected by crystalline symmetry {\color{blue}\cite{7,8,9,10,11,12,13,14}},
enabling spin-dependent transport without relying on spin--orbit coupling.
A representative example is the metal--organic framework (MOF) $\mathrm{[C(NH_2)_3]Cr(HCOO)_3}$  {\color{blue}\cite{18}},
which enables a homojunction tunnel junction with a predicted giant magnetoresistance of up to 53\%.
Recently, numerous candidate altermagnets have been identified,
including MnTe {\color{blue}\cite{1, 20}}, KV$_2$Se$_2$O {\color{blue}\cite{21}},
RuO$_2$ {\color{blue}\cite{2}}, CrSb {\color{blue}\cite{23}}, Rb$_{1-\delta}$V$_2$Te$_2$O {\color{blue}\cite{24}},
and RuF$_4$ {\color{blue}\cite{25}},
highlighting the broad chemical and structural diversity of this emerging material class. Their compensated magnetic order and momentum-dependent spin splitting make altermagnets promising platforms for low-dissipation and high-efficiency spintronic devices.

Owing to these distinctive advantages, developing effective routes to realize altermagnetism has become a central research goal.
For example, a variety of theoretical strategies have been proposed, including building spin-lattice models of altermagnetism~{\color{blue}\cite{26_1,26}}, stacking engineering of two-dimensional (2D) materials~{\color{blue}\cite{27}}, MOF-based inverse design~{\color{blue}\cite{28,34_5,29,30,32}}, high-throughput screening~{\color{blue}\cite{33,34}}, and chemical modification (e.g., ion intercalation)~{\color{blue}\cite{34_1}}.
Nevertheless, experimentally realized altermagnetic materials remain scarce~{\color{blue}\cite{song}}, making the development of simple and experimentally feasible design strategies highly desirable.
Moreover, achieving 100\% spin polarization along certain directions around Fermi level for altermagnets remains challenging. However, no altermagnetic system reported to date has been demonstrated to exhibit single-spin-channel transport.
Consequently, it is highly desirable to systematically search for and design altermagnetic systems with full spin polarization along certain directions.

In this article, we propose a chemical strategy for realizing 100\% spin-polarized transport in 2D altermagnetic organometallics by incorporating ambidentate ligands into square-lattice frameworks.
Without loss of generality, we consider an altermagnet with spin splitting along two representative reciprocal-space directions, $k_1$ and $k_2$,  with the corresponding real-space transport directions denoted by $x_1$ and $x_2$, respectively.
As illustrated in Figure~1(a), in a conventional altermagnet, both spin-up and spin-down bands are dispersive along $k_1$.
Consequently, both spin channels possess finite group velocities ($v_{1}^{\uparrow},v_{1}^{\downarrow}\neq0$) and contribute to the conductivity along the corresponding $x_1$ direction ($\sigma_{1}^{\uparrow},\sigma_{1}^{\downarrow}>0$), resulting in only partial spin polarization [Figure~1(b)].
A similar situation occurs along $k_2$ and the corresponding $x_2$ direction.
To maximize the conductivity difference between the two spin channels, a natural idea  is to design a new altermagnet in which the spin-split bands consist of a highly dispersive band in one spin channel and a completely flat band in the opposite spin channel, as schematically shown in Figure~1(c).
For example, along $k_1$, if the spin-up band remains dispersive while the spin-down band becomes flat, the spin-up channel retains a finite group velocity and conductivity ($v_{1}^{\uparrow}\neq0$, $\sigma_{1}^{\uparrow}>0$), whereas the spin-down channel has a vanishing group velocity and conductivity ($v_{1}^{\downarrow}=0$, $\sigma_{1}^{\downarrow}=0$).
The resulting charge transport along $x_1$ is therefore carried exclusively by the spin-up channel, giving rise to 100\% spin polarization [Figure~1(d)].
Conversely, the spin-down channel dominates the transport along $x_2$, leading to a spin polarization of $-100\%$.
These considerations suggest that the coexistence of altermagnetically spin-split dispersive and flat bands provides a direct route to fully spin-polarized and direction-dependent transport.
Therefore, realizing 100\% spin polarization requires identifying altermagnetic systems in which dispersive and flat bands coexist in opposite spin channels.
Motivated by this design principle, we further perform a tight-binding analysis of ambidentate-ligand based square lattices and demonstrate that such a band configuration can indeed be realized.
Finally, we screen 90 candidate MOFs and identify Fe(taz)$_2$Cl$_2$, Co(app)$_2$Cl$_2$, and Fe(app)$_2$Cl$_2$ as representative altermagnets exhibiting 100\% spin polarization along the [110] direction together with pronounced piezomagnetic responses.

\begin{figure}[!tbp]
\includegraphics[width=1\linewidth]{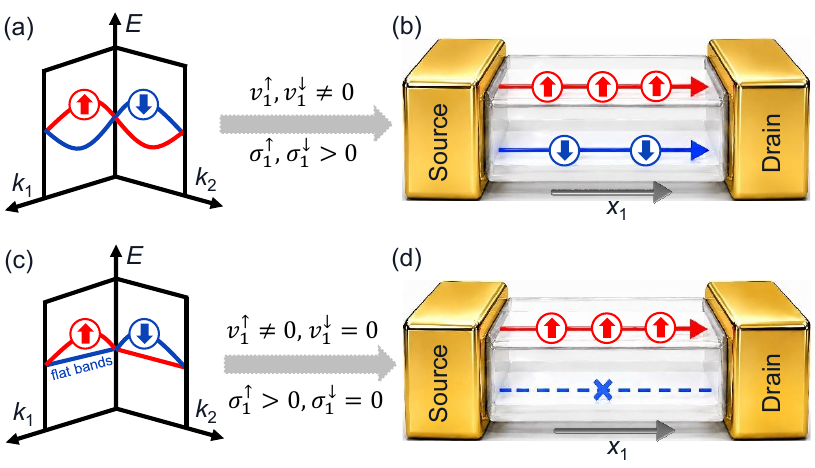}
\caption{Schematic illustration of spin-dependent transport in altermagnets.
(a) A conventional altermagnet with $k$-dependent spin splitting along two arbitrary reciprocal-space directions, $k_1$ and $k_2$.  Both spin-up and spin-down bands are dispersive, giving rise to finite group velocities ($v_{1}^{\uparrow},v_{1}^{\downarrow}\neq0$) and conductivities ($\sigma_{1}^{\uparrow},\sigma_{1}^{\downarrow}>0$) along the $x_1$ direction, where $x_1$ denotes the real-space direction corresponding to $k_1$.
(b) Corresponding transport schematic along the $x_1$ direction, with red and blue channels representing spin-up and spin-down conduction, respectively.
(c) A new altermagnet with coexisting spin-split dispersive and flat bands. Along the $k_1$ direction, the dispersive spin-up band exhibits a finite group velocity ($v_{1}^{\uparrow}\neq0$), resulting in a finite conductivity along the corresponding real-space $x_1$ direction ($\sigma_{1}^{\uparrow}>0$), whereas the flat spin-down band has a vanishing group velocity along $k_1$ ($v_{1}^{\downarrow}=0$) and hence zero conductivity along $x_1$ ($\sigma_{1}^{\downarrow}=0$).
(d) Resulting spin-selective transport, with a conductive spin-up channel and a suppressed spin-down channel, giving rise to 100\% spin polarization along the $x_1$ direction.}
\end{figure}

{\bf{2. Results}}

{\bf{2.1 Tight-binding model analysis.}}

Ambidentate ligands  ~{\color{blue}\cite{ab}} are widely used in coordination chemistry and contain two different donor atoms that can selectively coordinate to transition-metal (TM) atoms.
To investigate altermagnetism in a 2D square lattice formed by the ambidentate ligands and the TM atoms, we construct the tight-binding (TB) model shown in Figures 2 and S1 of Section A in the Supplemental Material (SM).
 The parent square lattice contains two magnetic sublattices, $\mathrm{TM}_{\uparrow}$ and $\mathrm{TM}_{\downarrow}$,
 whose opposite spin moments are compensated by spin-space group symmetries in the absence of spin-orbit coupling [Figure S1(a)].
 In an ideal symmetric lattice without ambidentate ligands, the two TM sites possess equivalent $D_4$ symmetries and are connected by real-space operations such as rotation, inversion, and mirror symmetry combined with a spin-space $C_2$ operation. By contrast, the ambidentate ligands introduce opposite local distortions around the two TM sublattices,
 reducing the local symmetries from $D_4$ to $D_2$ and breaking the inversion-like symmetry that connects $\mathrm{TM}_{\uparrow}$ and $\mathrm{TM}_{\downarrow}$ [Figures S1(b) and S1(c)].
 This ambidentate-ligand-induced symmetry reduction makes the two spin sublattices locally inequivalent while maintaining compensated magnetic order, thereby giving rise to altermagnetic spin splitting along the M-$\Gamma$--M$^\prime$ paths in the 2D Brillouin zone (BZ).
As schematically illustrated in Figure 2(a), the resulting altermagnetic state is characterized by opposite spins occupying neighboring TM sites,
where the red and blue arrows denote the spin-up and spin-down states of TMs,
respectively, and the gradient-colored trapezoids represent the ambidentate ligands.
To determine the role of different parameters, such as electronic hopping and exchange splitting, in the strength of the altermagnetic spin splitting, we next carry out a low-energy TB analysis. %
Accordingly, a minimal four-band TB model is constructed in the Bloch function basis, ordered as \(\{\mathrm{TM}_{\uparrow}-d_{xz}, \mathrm{TM}_{\uparrow}-d_{yz}, \mathrm{TM}_{\downarrow} -d_{xz}, \mathrm{TM}_{\downarrow}-d_{yz}\}\). %
The Hamiltonian is written as: \begin{equation} \mathcal{H}_t = \alpha(\vec{k}) \mathbb{I} + \beta(\vec{k})\Sigma_z \otimes \sigma_x + \gamma(\vec{k})\Sigma_x \otimes \sigma_0 + \delta(\vec{k})\Sigma_x \otimes \sigma_x, \tag{1} \end{equation} where \(\mathbb{I}\) is a \(4 \times 4\) identity matrix. \(\vec{\Sigma}\) and \(\vec{\sigma}\) are the Pauli matrices in the sublattice bases formed by $\mathrm{TM}_{\uparrow}$ and $\mathrm{TM}_{\downarrow}$, and in the orbital basis composed of \(d_{xz}\) and \(d_{yz}\), respectively, while \(\sigma_0\) is the identity matrix.
\begin{figure}[!tbp]
\includegraphics[width=1\linewidth]{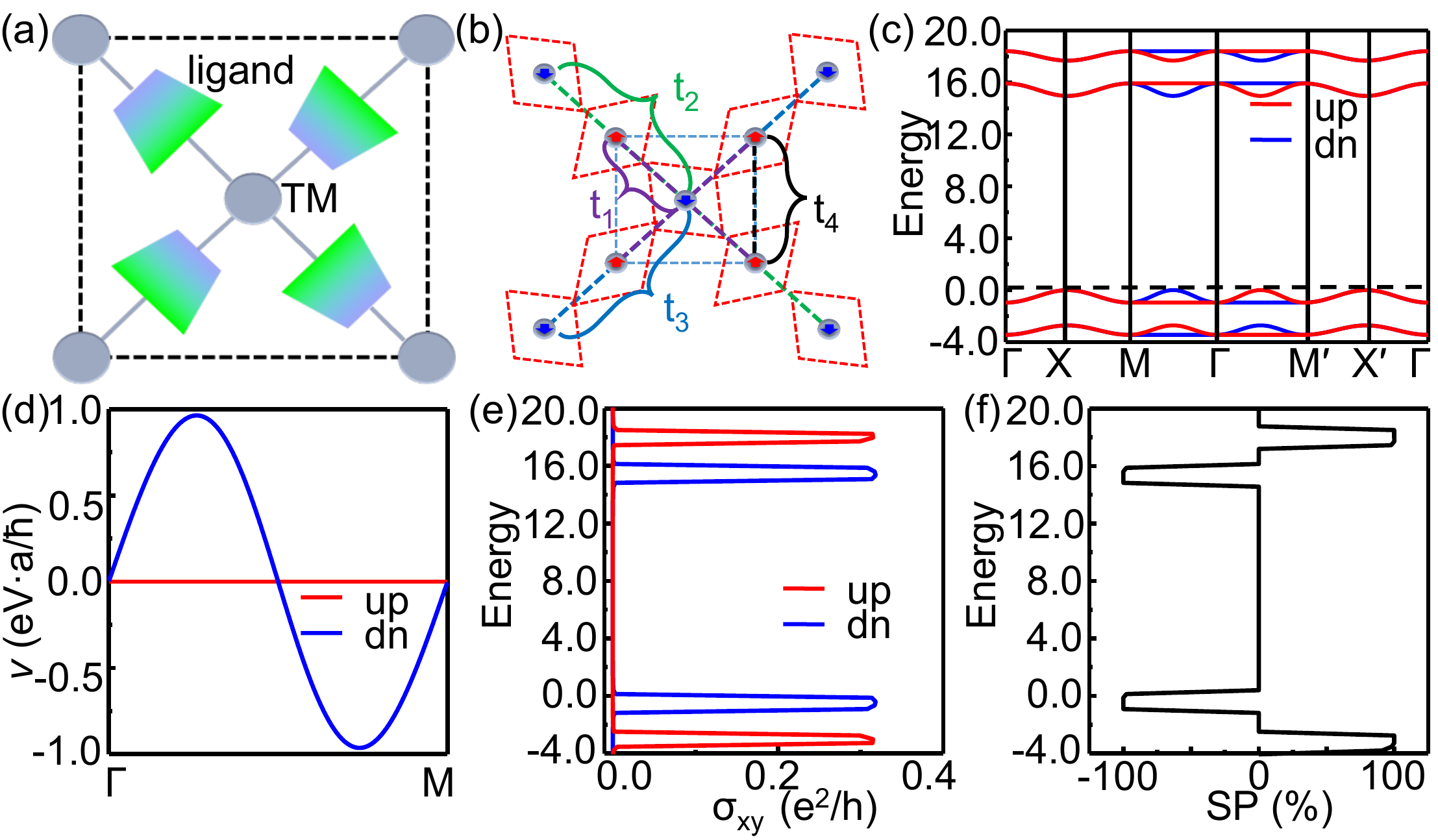}
\caption{Tight-binding model, band structures, and spin-transport properties. (a) Altermagnetic state in a 2D square lattice containing ambidentate ligands, with opposite spins located at lower-symmetry TM sites.
Red and blue arrows indicate spin-up and spin-down, respectively, and gradient-colored trapezoids denote the ambidentate ligands. (b) Hopping processes on the square lattice: $t_1$ (purple line),
nearest-neighbor hopping between opposite-spin sites; $t_2$ (royal blue line) and $t_3$ (green line), next-nearest-neighbor hoppings between same-spin sites; $t_4$ (black line),
nearest-neighbor hopping between same-spin sites. The ambidentate ligands give rise to $t_2$ = $t_3$. (c) Tight-binding band structure with $t_1$ = 0.036, $t_2$ = 0.036, $t_3$ = 0.036, $t_4$ = -0.60, $\xi_1$ = -0.55, $\xi_2$ = -0.55, and $\Delta_s$ = 0.10.
In (d) and (e), the red and blue curves represent the spin-up and spin-down channels, respectively, while the black curve in (f) denotes the spin polarization.}
\end{figure}
The functions \(\alpha(\vec{k})\), \(\beta(\vec{k})\), \(\gamma(\vec{k})\), and \(\delta(\vec{k})\) are determined by the on-site energies $\xi_n$ $(n=1, 2)$ and the effective $d-d$ hopping parameters $t_i$ ($i=1-4$) illustrated in Figure 2(b).
For simplicity, only hopping processes up to the second-nearest neighbors are considered, with \(t_1\) representing the nearest-neighbor hopping between opposite-spin sites, \(t_4\) the nearest-neighbor hopping between same-spin sites, and \(t_2\) and \(t_3\) the next-nearest-neighbor hoppings between same-spin sites. The condition $\left|t_2| = | t_3 \right|$ arises from the asymmetry introduced by ambidentate ligands.
The explicit expressions of these functions are
\begin{equation}
\begin{aligned}
\alpha(\vec{k}) &= \xi_1 + 2t_1 \cos(k_z c),\\
\beta(\vec{k}) &= \xi_2 + 2t_4 \cos(k_z c),\\
\gamma(\vec{k}) &= 8t_2 \cos\left(\frac{k_x a}{2}\right) \cos\left(\frac{k_y b}{2}\right) \cos\left(\frac{k_z c}{2}\right),\\
\delta(\vec{k}) &= -8t_3 \sin\left(\frac{k_x a}{2}\right) \sin\left(\frac{k_y b}{2}\right) \cos\left(\frac{k_z c}{2}\right).
\end{aligned}
\tag{2}
\end{equation}
Here, \(a\), \(b\), and \(c\) are the lattice constants of the tetragonal structure, and $k_z=0$ for the 2D case.

Diagonalizing the Hamiltonian in Eq. (1), with the functions defined in Eq. (2), yields four energy eigenvalues:
\begin{equation}
\begin{aligned}
\mathcal{E}_{\pm}(\vec{k}) &= \alpha(\vec{k}) - \left\{\beta(\vec{k})^2 + [\delta(\vec{k}) \pm \gamma(\vec{k})]^2\right\}^{1/2},\\
\mathcal{E}_{\pm}^{\dagger}(\vec{k}) &= \alpha(\vec{k}) + \left\{\beta(\vec{k})^2 + [\delta(\vec{k}) \pm \gamma(\vec{k})]^2\right\}^{1/2}.
\end{aligned}
\tag{3}
\end{equation}
The corresponding eigenvectors indicate that there is an energy splitting \(\Delta\mathcal{E} = \mathcal{E}_{+}(\vec{k}) - \mathcal{E}_{-}(\vec{k})\) between bands with dominant $\mathrm{TM}_{\uparrow}$ and $\mathrm{TM}_{\downarrow}$ sublattice contributions.
Note that such an energy splitting between bands of different sublattice characters is present prior to including the effect of altermagnetic exchange splitting \(\Delta_s\). We now show that the inclusion of \(\Delta_s\) translates the sublattice splitting into the spin splitting of the bands.

To include the effect of the altermagnetic exchange, we rewrite Eq. (1) in the basis of
\(\{\mathrm{TM}_{\uparrow}{-}d_{xz,\uparrow},\ \mathrm{TM}_{\uparrow}{-}d_{yz,\uparrow},\
\mathrm{TM}_{\downarrow}{-}d_{xz,\uparrow},\ \mathrm{TM}_{\downarrow}{-}d_{yz,\uparrow},\\
\mathrm{TM}_{\uparrow}{-}d_{xz,\downarrow},\ \mathrm{TM}_{\uparrow}{-}d_{yz,\downarrow},\
\mathrm{TM}_{\downarrow}{-}d_{xz,\downarrow},\ \mathrm{TM}_{\downarrow}{-}d_{yz,\downarrow}\}\),
and add the exchange term \(\mathcal{H}_{\text{eff}}^{\text{TM}} = \Delta_s\mathcal{S}_z \otimes (\Sigma_z \otimes \sigma_0)\). The full Hamiltonian is
\begin{equation}
\mathcal{H} = S_0 \otimes \mathcal{H}_t + \Delta_s\mathcal{S}_z \otimes (\Sigma_z \otimes \sigma_0). \tag{4}
\end{equation}
Here, \(S_z\) and \(S_0\) are the Pauli matrices and the identity matrix in the spin basis, respectively. Therefore, the eigenvalue equation can be written as $E_{{\uparrow\downarrow}}^{\pm} = \alpha(\vec{k}) \pm \sqrt{(\pm\Delta_s\ \pm \beta(\vec{k}))^2 + (\gamma(\vec{k}) \pm \delta(\vec{k}))^2}$.
The band gap of this model can be expressed as $\Delta_{\mathrm{gap}} = 2(\xi_2 + 2t_4) - \Delta_s$.
A positive $\Delta_{\mathrm{gap}}$ corresponds to a semiconducting state. Metallic behavior occurs when \(\xi_2 + 2t_4 \le \Delta_s/2\).
Along the M-$\Gamma$ and $\Gamma$--M$^\prime$ paths, each band splits into a dispersive band and a flat band when $\left|t_2| = | t_3 \right|$.
In addition, this Hamiltonian describes conventional antiferromagnetic states for \(\Delta_s = 0\) and altermagnetic states for \(\Delta_s \neq 0\).
The exchange splitting energy between the spin-up and spin-down bands is extracted from the computed spin-resolved densities of states of this TB model.
Using the parameters $t_1$ = 0.036, $t_2$ = 0.036, $t_3$ = 0.036, $t_4$ = -0.60, $\xi_1$ = -0.55, $\xi_2$ = -0.55, and $\Delta_s$ = 0.10, we calculate the TB band structures along the high-symmetry paths $\Gamma$--X--M--$\Gamma$--M$^\prime$--X$^\prime$--$\Gamma$, as shown in Figure 2(c).
The calculated bands exhibit the expected altermagnetic spin splitting along the $\Gamma$--M and $\Gamma$--M$^\prime$ paths, with one spin channel forming flat bands.
Since the electronic dispersion in a periodic lattice is described by $E(k)$, the corresponding group velocity is given by \( v = \frac{1}{\hbar} \frac{dE}{dk} \).
For the flat band along the $\Gamma$--M path, \( dE_{\uparrow}/dk = 0 \), and thus the electron group velocity in the spin-up channel becomes zero {\color{blue}\cite{34_0}} in the valence-band maximum [Figure 2(d)], strongly suppressing long-range electron propagation in this 2D square lattice.
Consequently, this spin-split flat band makes a negligible contribution to the transverse conductivity $(\sigma_{xy})$, whereas the other spin channel remains conductive and dominates $\sigma_{xy}$ [Figure 2(e)], thereby achieving 100\% spin polarization along the [110] direction [Figure 2(f)].

\begin{figure}[!tbp]
\includegraphics[width=1\linewidth]{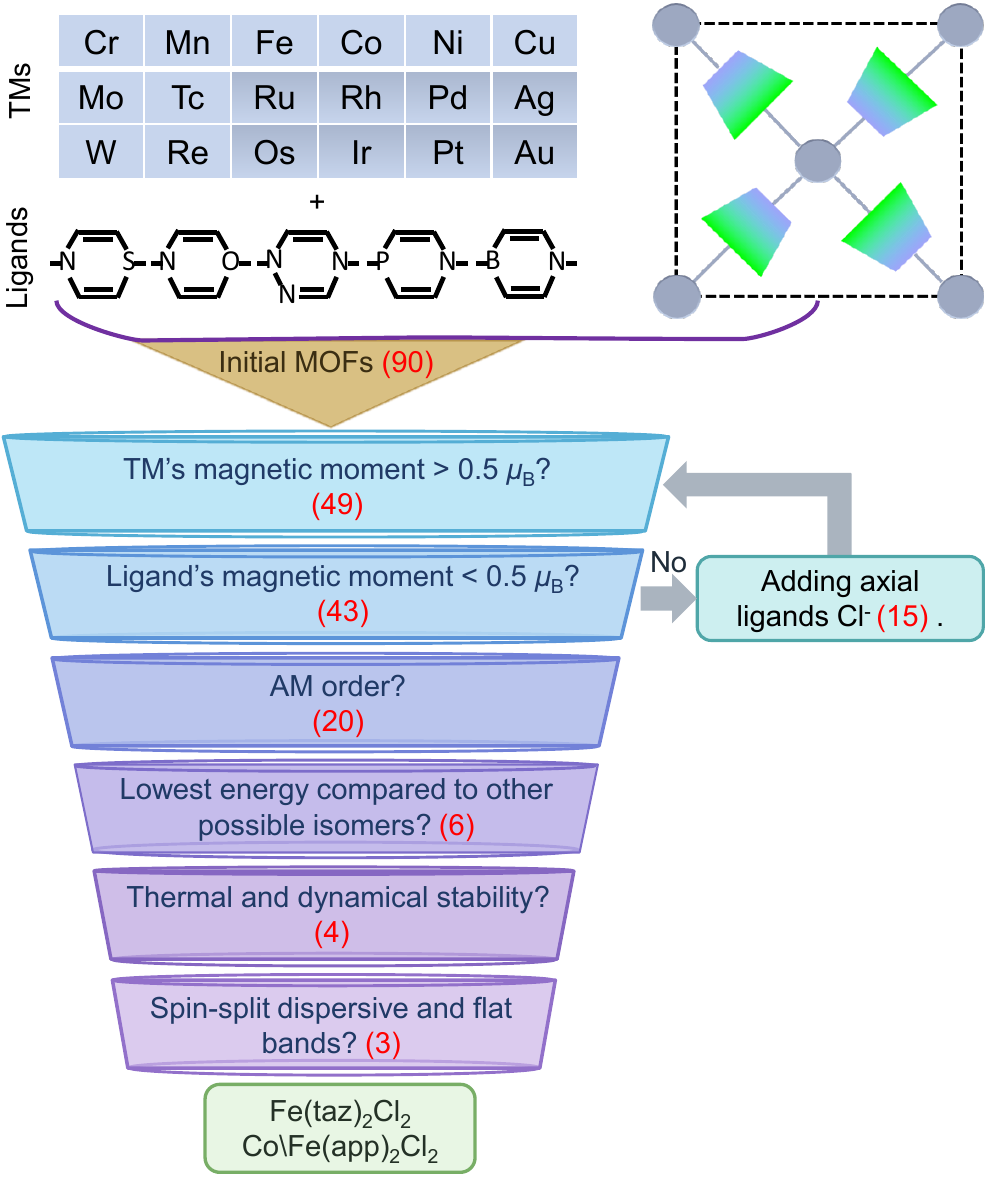}
\caption{Screening workflow for 2D altermagnetic (AM) MOFs. Based on 18 TM elements and 5 kinds of ambidentate ligands, we construct 90 two-dimensional MOFs and screen out three altermagnetic candidates: Fe(taz)$_2$Cl$_2$, Co(app)$_2$Cl$_2$, and Fe(app)$_2$Cl$_2$. }
\end{figure}

{\bf{2.2 Systematic search for 2D altermagnetic MOFs.}}

Referring to the defined structure shown in Figure 2(a), we combine 18 TM elements with 5 organic ligands to construct 2D MOF structures with square-lattice characteristics (Figure 3), yielding 90 initial candidate systems. A stepwise screening procedure is then performed.
Firstly, the local magnetic moments of the TM atoms are examined, and systems with a TM magnetic moment greater than $0.5~\mu_{\mathrm{B}}$ are screened to ensure that the TM sites could serve as potential magnetic sites for altermagnetism, reducing the number of candidates to 49.
Next, the magnetic moments on the ligands are evaluated. The magnetic moment of a ligand in these systems should be less than $0.5~\mu_{\mathrm{B}}$, so as to prevent the formation of ferrimagnetic order induced by the $d--p$ spin coupling between the ligands and the TM atoms {\color{blue}\cite{34_11,34_12,34_13,34_14,34_15}}.
For systems with a ligand magnetic moment larger than 0.5 $\mu_{\mathrm{B}}$,  axial inorganic Cl$^-$ ligands are introduced to reduce the metal-ligand charge transfer (15 candidates), after which the magnetic-moment criteria are reassessed. As a result, we obtain 43 candidate materials with small ligand magnetic moments.

Subsequently, the structures with an altermagnetic ground state are identified, yielding 20 candidates.
In addition, three other possible isomers exist within the same square lattice framework (Figure S2 in Section A of the SM). By comparing the ground-state energies of the four isomers, we identify 6 systems in which the altermagnetic isomer has the lowest energy. The metastable altermagnetic MOFs are presented in Section B of the SM.
Then, the remaining candidates are subjected to dynamical and thermal stability tests, leaving 4 candidates. Finally, we investigate the band-structure requirements of the altermagnetic systems for spin-split dispersive and flat bands, ensuring that 100\% spin polarization can be achieved along the [110] direction, leading to the identification of 3 promising 2D altermagnetic candidates, as shown in Figures 4(a1)--4(c1): Fe(taz)$_2$Cl$_2$, Co(app)$_2$Cl$_2$, and Fe(app)$_2$Cl$_2$. Their detailed lattice-structure information is provided in Tables VI--VIII in Section C of the SM.

{\bf{2.3 Stability, magnetism, and spin-polarized transport}}

Having identified the 3 promising candidates, we next discuss their structural stability in detail.
As shown in Figures S4(a), S4(d), and S4(g) in Section D of the SM, the phonon spectra exhibit no imaginary frequencies throughout the 2D BZ, confirming their dynamical stability.
The weak dispersion of the phonon bands originates from the large lattice parameters typical of MOFs, while the soft phonon modes reflect their structural flexibility {\color{blue}\cite{34_2}}.
The results of 6 $ps$ $ab$ initio molecular dynamics (AIMD) simulations at 200 and 300 K are shown in Figure S4 in Section D of the SM.
At 300 K, Fe(taz)$_2$Cl$_2$ maintains a stable total-energy profile with only minor thermal fluctuations and no abrupt structural disruption, indicating its room-temperature thermal stability.
For Co(app)$_2$Cl$_2$ and Fe(app)$_2$Cl$_2$, structural integrity is retained at 200 K, suggesting that these two candidates are thermally stable under relatively low-temperature conditions.


After discussing the stability of these systems, we investigate the magnetic properties.
As a representative example, Figure 4(a1) illustrates the spin-density distribution of Fe(taz)$_2$Cl$_2$ in its altermagnetic ground state.
The magnetic moments are mainly localized on the Fe atoms, with neighboring Fe sites exhibiting antiparallel coupling, whereas the organic ligands contribute negligibly.
The local magnetic moment on each Fe atom is about 3.76 $\mu_{\mathrm{B}}$, corresponding to a formal spin of $S = 2$. Similarly, the spin-density distributions of Co(app)$_2$Cl$_2$ and Fe(app)$_2$Cl$_2$ are presented in Figures 4(b1)--4(c1), showing features very similar to those of Fe(taz)$_2$Cl$_2$.

To estimate the N\'{e}el temperature  ($T\rm_N$) of the three altermagnetic MOFs, Monte Carlo simulations are performed on the basis of the Heisenberg model with the Hamiltonian given by
\begin{equation}
	\begin{aligned}
		H = -\sum_{i,j} J S_i \cdot S_j + \sum_i D_i S_{iz}^2
	\end{aligned}
	\tag{5}
\end{equation}
where $J$ denotes the nearest-neighbor exchange parameter of the TM atoms, as illustrated in Figure 4(a1), with values of -0.34, -0.33, and -0.55 meV for Fe(taz)$_2$Cl$_2$, Co(app)$_2$Cl$_2$, and Fe(app)$_2$Cl$_2$, respectively.
$S_i$ and $D_i$ represent the spin and the magnetic anisotropy parameter of the TM atoms, respectively.
The corresponding $D_i$ values for TMs are 1.42, 1.91, and 9.46 meV for Fe(taz)$_2$Cl$_2$, Co(app)$_2$Cl$_2$, and Fe(app)$_2$Cl$_2$, respectively.

Figure S6(a) in Section E of the SM presents the temperature-dependent specific heat \( C_v \) of the three systems, where \( C_v = \frac{\langle E^2 \rangle - \langle E \rangle^2}{k_B T^2} \).
The $T_{\rm N}$, defined by the peak position of the specific heat curve \( C_v \), is estimated to be 5.0, 7.5, and 8.5~K for Fe(taz)$_2$Cl$_2$, Co(app)$_2$Cl$_2$, and Fe(app)$_2$Cl$_2$,
respectively---values that are lower than those of V$_2$Se$_2$O ($T\rm_N$ = 600 K)~{\color{blue}\cite{45}} and Fe$_2$Se$_2$O ($T\rm_N$ = 319 K)~{\color{blue}\cite{46}},
while remaining comparable to those reported in 2D altermagnetic MOFs (Ca(pyz)$_2$: $T\rm_N$ = 15.5 K, Sr(pyz)$_2$: $T\rm_N$ = 11.5 K)~{\color{blue}\cite{34_5}}.
Notably, the application of uniaxial strain along the [110] direction can significantly enhance the magnetic transition temperature.
For example, under $-6\%$ strain, the \( T\rm_ N \) of Fe(taz)$_2$Cl$_2$ increases to 43~K, as shown in Figure S6(b) in Section E of the SM.

Besides enhancing \( T\rm_ N \), uniaxial strain along the [110] direction breaks the inversion symmetry at the magnetic centers and thus induces piezomagnetism.
As shown in Figures S6(c)--S6(f) in Section E of the SM, all systems acquire finite net magnetic moments within the strain range of \(-6\%\) to \(+6\%\), with a stronger response under tensile strain.
Specifically, at \(+6\%\) tensile strain, the net magnetic moments of Co(app)$_2$Cl$_2$ and Fe(taz)$_2$Cl$_2$ reach approximately \(-0.09\) and \(0.01\,\mu\rm_B\), respectively, whereas at \(+6\%\) tensile strain, that of Fe(app)$_2$Cl$_2$ is approximately 0.02 $\mu_{\mathrm{B}}$.
The piezomagnetic response is further strengthened by the introduction of axial Cl$^-$ ligands.
In particular, the \(-0.09\,\mu\rm_B\) obtained for Co(app)$_2$Cl$_2$ is already comparable to the values reported for inorganic monolayers such as V$_2$Se$_2$O and Fe$_2$Se$_2$O {\color{blue}\cite{45,46}}.

\begin{figure}[!tbp]
\includegraphics[width=1\linewidth]{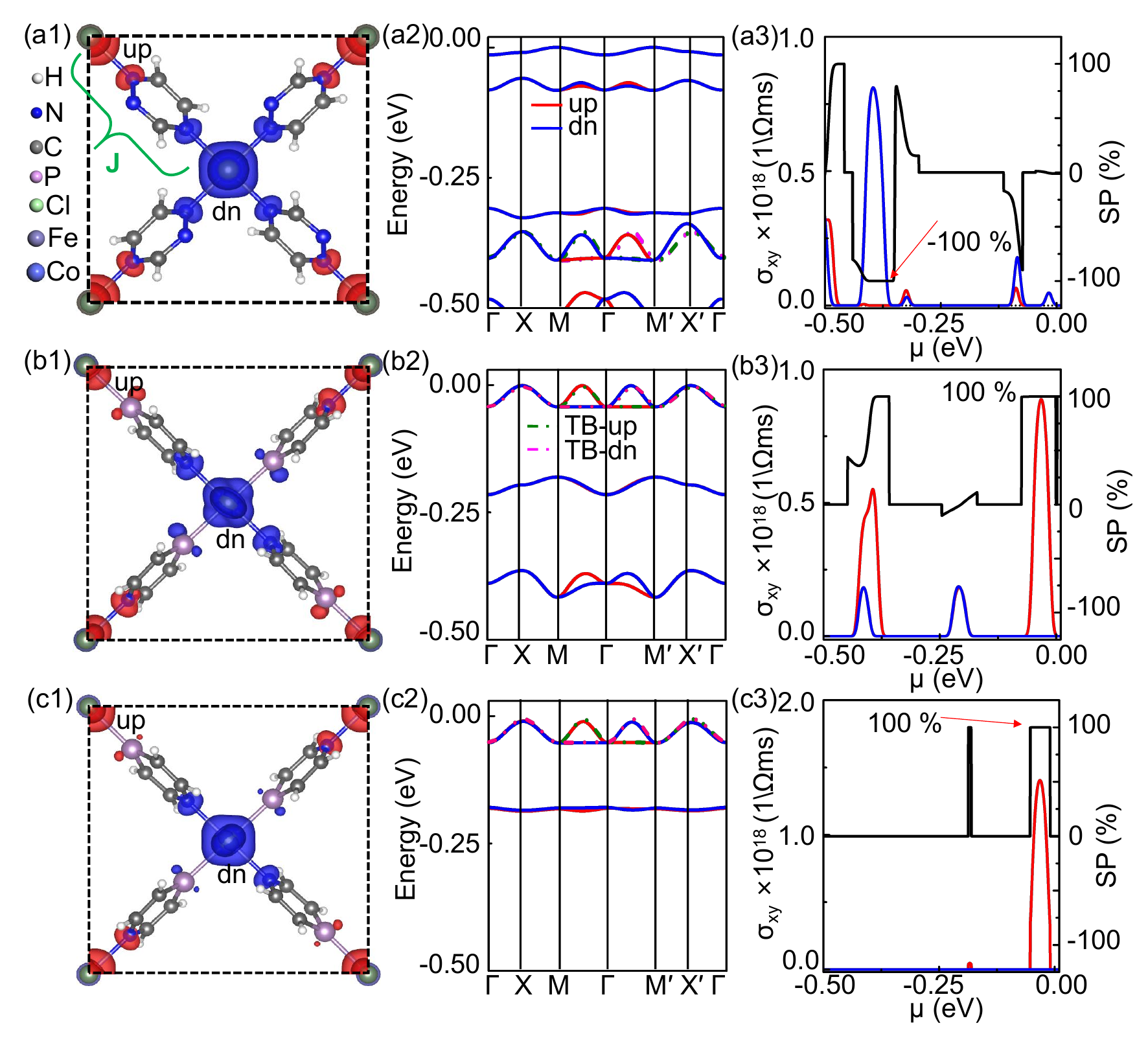}
\caption{(a1)-(c1) Crystal structures and spin-density distribution with an isovalue of 0.003 $\text{\AA}^{-3}$ of Fe(taz)$_2$Cl$_2$, Co(app)$_2$Cl$_2$, and Fe(app)$_2$Cl$_2$, where red and blue isosurfaces denote the spin-up and spin-down components, respectively; 	$J$ labels the exchange interaction between neighboring TMs. White, blue, black,  purple, green, gray, and royal blue spheres denote H, N, C, P, Cl, Fe, and Co. (a2)-(c2) Band structures of Fe(taz)$_2$Cl$_2$, Co(app)$_2$Cl$_2$, and Fe(app)$_2$Cl$_2$ below the Fermi level. The red and blue solid lines represent the spin-up and spin-down bands calculated by DFT, respectively, while the green and pink dashed lines denote the corresponding bands fitted using the TB model. The Fermi level is set to the valence-band maximum. (a3)-(c3) The transverse conductivities $\sigma_{xy}$ and spin polarization (SP) of the three systems.}
\end{figure}


Having established the magnetic configurations, we also calculate the band structures of Fe(taz)$_2$Cl$_2$, Co(app)$_2$Cl$_2$, and Fe(app)$_2$Cl$_2$ using the PBE+U functional to characterize their altermagnetic electronic features, as shown in Figures 4(a2)--4(c2) and Figure S7. Note that the HSE06 band structures (Figure S8)also reveal semiconducting behavior for all three systems, consistent with the PBE+U results presented in Section F of the SM (Figures S7). Along the $\Gamma$--$\rm X$--$\rm M$ and $\rm M'$--$\rm X'$--$\Gamma$ paths, no spin splitting is observed because the spin-space symmetry $\{C_{2z}\,||\,C_{4z}\}$ enforces spin degeneracy along these directions. Conversely, spin splitting appears along the $\rm M$--$\Gamma$--$\rm M'$ path protected by  $\{C_{2z}\,||\,C_{4z}\}$ symmetry, where red and blue lines represent the spin-up and spin-down channels, respectively. Across the entire 2D BZ, the spin-up and spin-down bands are fully compensated, resulting in a zero macroscopic magnetic moment. Notably, partial flat bands emerge along the $\rm M$--$\Gamma$--$\rm M'$ paths in Figures~4(b2)--4(c2), indicating localized electronic states and strongly suppressed transport in the corresponding spin channel, which ultimately leads to 100\% spin polarization along the [110] direction.

Furthermore, we fit the parameters of the TB model to the DFT band structures. The comparison between the DFT (red and blue solid lines) and fitted TB (green and pink dashed lines) band structures is shown in Figures 4(a2)--4(c2), demonstrating that the TB model captures the main low-energy band features of these systems.
The fitted parameters are as follows: Fe(taz)$_2$Cl$_2$: $t_1 = 0.62$, $t_2 = -0.037$, $t_3 = 0.037$, $t_4 = -5.31$, $\xi_1 = 9.99$, $\xi_2 = -3.17$, and $\Delta_s = 2.52$;
Co(app)$_2$Cl$_2$: $t_1 = 0.56$, $t_2 = 0.034$, $t_3 = 0.034$, $t_4 = -5.62$, $\xi_1 = 9.99$, $\xi_2 = -2.03$, and $\Delta_s = -1.15$;
 and Fe(app)$_2$Cl$_2$: $t_1 = 0.61$, $t_2 = 0.031$, $t_3 = 0.031$, $t_4 = -9.99$, $\xi_1 = 9.41$, $\xi_2 = 5.41$, and $\Delta_s = 3.96$. All fitted parameters are given in $eV$. The results indicate that 100\% spin polarization along the [110] direction emerges when the hopping amplitudes satisfy $|t_2| = |t_3|$. 
%

%
The spin-polarized transverse conductivity $\sigma_{xy}$ along the [110] direction, together with the corresponding spin polarization (black solid lines), is evaluated using the Boltzmann transport equation {\color{blue}\cite{47}}, as described in Section G of the SM.
The calculated results are shown in Figures 4(a3)--4(c3), where the red and blue solid lines represent the spin-up and spin-down channels, respectively.
Consistent with the altermagnetic band splitting along the [110] direction, the spin-resolved $\sigma_{xy}$ exhibits a pronounced channel asymmetry near the Fermi level.

For Fe(taz)$_2$Cl$_2$, the altermagnetically spin-split bands about $-0.34$ eV below the Fermi level consist of dispersive and nearly flat branches, as shown in Figure 4(a2). Consistent with the TB analysis, the nearly flat band contributes negligibly to $\sigma_{xy}$, whereas the dispersive band in the opposite spin channel dominates the electrical transport. Consequently, complete spin polarization is achieved within the energy window from $-0.41$ to $-0.38$ eV [Figure 4(a3)]. However, because this window lies relatively deep below the valence-band maximum, accessing it may require substantial hole doping, which limits its practical applicability.

\begin{table}[!tbp]
	\centering
	\caption{Summary of spin polarization in altermagnets.}
	\begin{tabular}{@{}lc@{}}
		\toprule
		\textbf{Materials} & \textbf{Spin Polarization} \\
		\midrule
		Mn$_2$PSe$_3$ {\color{blue}\cite{mn2pse3}} & 58.2\% \\
		Ca(pyz)$_2$ {\color{blue}\cite{34_5}} & 56.85\% \\
		Sr(pyz)$_2$ {\color{blue}\cite{34_5}} & 86.04\% \\
		RuO$_2$ {\color{blue}\cite{ruo2}} & 70\% \\
		Cr$_2$O (monolayer) {\color{blue}\cite{cr2o}} &  95.2\% \\
		CrS bilayer {\color{blue}\cite{crs}} & 87\% \\
		KV$_2$Se$_2$O {\color{blue}\cite{zhang}} & 60\% \\
		\bottomrule
	\end{tabular}
\end{table}

In contrast, for Co(app)$_2$Cl$_2$ and Fe(app)$_2$Cl$_2$, the altermagnetically spin-split dispersive and nearly flat bands occur near the valence-band maxima [Figures 4(b2)--4(c2)]. As a result, 100\% spin polarization along the [110] direction is achieved within the energy window of $-0.04$-$0.00$ eV for Co(app)$_2$Cl$_2$ and $-0.05$-$0.00$ eV for Fe(app)$_2$Cl$_2$ near the Fermi level [Figures 4(b3)--4(c3)], exceeding the spin polarization values previously reported for other altermagnetic materials, as summarized in Table I. Because these two fully spin-polarized energy windows are located close to the valence-band maxima, they may be experimentally accessed through moderate p-type doping, making these two materials promising for spin-polarized transport applications.

{\bf{3. Summary}}

In summary, we propose a chemical strategy for realizing 100\% spin-polarized transport along the [110] direction in 2D altermagnetic organometallic lattices by incorporating ambidentate ligands into square-lattice frameworks.
Tight-binding modeling reveals that the ligand-induced coordination environment breaks the equivalence of spin-dependent hopping channels, giving rise to altermagnetic spin splitting. More importantly, the coexistence of altermagnetically spin-split dispersive and flat bands near the valence- or conduction-band edges suppresses charge transport in one spin channel, thereby enabling fully spin-polarized conduction in the other.
Guided by this design principle, we systematically screened  90 two-dimensional MOFs and identified three promising altermagnetic candidates, namely Fe(taz)$_2$Cl$_2$, Co(app)$_2$Cl$_2$, and Fe(app)$_2$Cl$_2$.
First-principles calculations confirm that the candidate materials exhibit robust altermagnetic spin splitting and notable piezomagnetic responses, with all three exhibiting 100\% spin polarization within selected valence-band energy windows; notably, the fully spin-polarized windows of Co(app)$_2$Cl$_2$ and Fe(app)$_2$Cl$_2$ lie close to the valence-band maximum.
Our findings establish coordination-environment engineering as an effective route to designing highly spin-polarized altermagnets and provide a promising material platform for realizing single-spin-channel transport without net magnetization.

{\bf{4. Computational methods}}

\noindent
Based on density functional theory (DFT) implemented in the Vienna Ab Initio Simulation Package (\textsc{VASP}) {\color{blue}\cite{35,36}}, first-principles calculations are performed using the Perdew--Burke--Ernzerhof (PBE) formulation of the generalized gradient approximation (GGA). To properly account for the $3d$ electrons of the Fe atom, the GGA$+U$ method {\color{blue}\cite{37}} is employed, with the Coulomb interaction parameter $U$ set to 5~eV {\color{blue}\cite{38}} (U is set to 4 eV for Co {\color{blue}\cite{40}}) and the exchange interaction parameter $J$ set to 1~eV. A vacuum spacing of 20~\AA\ is introduced to minimize interactions between adjacent layers. The DFT-D3 method with Becke--Johnson damping is used to describe van der Waals (vdW) interactions between neighboring organic rings. The projected augmented wave (PAW) method is adopted, with a plane-wave cutoff energy of 520~eV. Structural optimization is considered converged when the total energy and atomic forces are below $10^{-6}$~eV and 0.01~eV/\AA, respectively. A $3\times 3\times 1$ Monkhorst--Pack $k$-point mesh is applied to sample the first Brillouin zone. For more accurate electronic structure results, hybrid functional calculations (HSE06) {\color{blue}\cite{41,42}} are carried out. AIMD simulations are performed for 6~$ps$ with a time step of 1~$fs$ in a $2\times 2\times 1$ supercell at 200 and 300~K, using the NVT ensemble with the Nos\'e--Hoover thermostat to control temperature. To examine dynamical stability, phonon spectra are calculated using the phonopy code {\color{blue}\cite{43}}. Monte Carlo simulations are performed using the Property Analysis and Simulation Package for Materials (PASP) {\color{blue}\cite{44}}.

\noindent
{\bf{Acknowledgements}}\\
\noindent

We gratefully acknowledge the National Natural Science Foundation of China (12404085, 22288201, 22573109, 22273092, and 22322304), the Innovation Program for Quantum Science and Technology (2021ZD0303306), USTC Tang Scholar, the computational resources are provided by the Supercomputing Center of USTC, and by the Shiyan Key Laboratory of Quantum Information and Precision Optics (Hubei University of Automotive Technology) No. SYZDK12024B03.\\

\bibliography{references}

\end{document}